\documentclass{emulateapj}
\usepackage{subfigure}

\def\lsim{\mathrel{\rlap{\lower4pt\hbox{\hskip1pt$\sim$}}
    \raise1pt\hbox{$<$}}}                % less than or approx. symbol
\def\gsim{\mathrel{\rlap{\lower4pt\hbox{\hskip1pt$\sim$}}
    \raise1pt\hbox{$>$}}}                % greater than or approx. symbol

\def\Kepler{\textit{Kepler}}

\shorttitle{All-Sky Gigapixel-Scale Telescopes}
\shortauthors{N.M. Law et al.}
\begin{document}
\title{Evryscope science: exploring the potential of all-sky gigapixel-scale telescopes}

\author{Nicholas M. Law\altaffilmark{1}, Octavi Fors\altaffilmark{1}, Jeffrey Ratzloff\altaffilmark{1}, Philip Wulfken\altaffilmark{1}, Dustin Kavanaugh\altaffilmark{1}, David J. Sitar\altaffilmark{2}, Zachary Pruett\altaffilmark{2}, Mariah N. Birchart\altaffilmark{2}, Brad N. Barlow\altaffilmark{3}, Kipp Cannon\altaffilmark{4}, S. Bradley Cenko\altaffilmark{5}, Bart Dunlap\altaffilmark{1}, Adam Kraus\altaffilmark{6}, Thomas J. Maccarone\altaffilmark{7}}

\altaffiltext{1}{Department of Physics and Astronomy, University of North Carolina at Chapel Hill, Chapel Hill, NC 27599-3255, USA}
\altaffiltext{2}{Department of Physics and Astronomy, Appalachian State University, Boone, NC 28608, USA}
\altaffiltext{3}{Department of Physics, High Point University, 833 Montlieu Avenue, High Point, NC 27268, USA}
\altaffiltext{4}{Canadian Institute for Theoretical Astrophysics, 60 St. George St., Toronto, ON, M5S 3H8}
\altaffiltext{5}{Goddard Space Flight Center, Mail Code 661, Greenbelt, MD 20771}
\altaffiltext{6}{Department of Astronomy, Univ. of Texas at Austin, 2515 Speedway, Stop C1400, Austin, TX 78712}
\altaffiltext{7}{Department of Physics, Texas Tech University, Box 40151, Lubbock TX 79409-1051}

\begin{abstract} Low-cost mass-produced sensors and optics have recently made it feasible to build telescope arrays which observe the entire accessible sky simultaneously. In this article we discuss the scientific motivation for these telescopes, including exoplanets, stellar variability and extragalactic transients. To provide a concrete example we detail the goals and expectations for the Evryscope, an under-construction 780 MPix telescope which covers 8,660 square degrees in each two-minute exposure; each night, 18,400 square degrees will be continuously observed for an average of $\approx$6 hours. Despite its small 61mm aperture, the system's large field of view provides an \'etendue which is $\sim$10\% of LSST. The Evryscope, which places 27 separate individual telescopes into a common mount which tracks the entire accessible sky with only one moving part, will return 1\%-precision, many-year-length, high-cadence light curves for every accessible star brighter than $\rm{\sim16^{th}}$ magnitude. The camera readout times are short enough to provide near-continuous observing, with a 97\% survey time efficiency. The array telescope will be capable of detecting transiting exoplanets around every solar-type star brighter than $\rm{m_V}$=12, providing at least few-millimagnitude photometric precision in long-term light curves. It will be capable of searching for transiting giant planets around the brightest and most nearby stars, where the planets are much easier to characterize; it will also search for small planets nearby M-dwarfs, for planetary occultations of white dwarfs, and will perform comprehensive nearby microlensing and eclipse-timing searches for exoplanets inaccessible to other planet-finding methods. The Evryscope will also provide comprehensive monitoring of outbursting young stars, white dwarf activity, and stellar activity of all types, along with finding a large sample of very-long-period M-dwarf eclipsing binaries. When relatively rare transients events occur, such as gamma-ray bursts (GRBs), nearby supernovae, or even gravitational wave detections from the Advanced LIGO/Virgo network, the array will return minute-by-minute light curves without needing pointing towards the event as it occurs. By co-adding images, the system will reach V$\sim$19 in one-hour integrations, enabling the monitoring of faint objects. Finally, by recording all data, the Evryscope will be able to provide pre-event imaging at two-minute cadence for bright transients and variable objects, enabling the first high-cadence searches for optical variability before, during and after all-sky events.
\end{abstract}


\maketitle

\section{Introduction}

Synoptic sky surveys generally cover very large sky areas to detect rare events. Since it is usually infeasible to cover thousands of square degrees with a single telescope, they repeatedly observe few-degree-wide fields, use large apertures to achieve deep imaging, and tile their observations across the sky. The resulting survey -- such as PTF \citep{Law2009}, Pan-STARRS \citep{Kaiser2010}, SkyMapper \citep{Keller2007}, CRTS \citep{Djorgovski2011}, ATLAS \citep{Tonry2011}, and many others -- is necessarily optimized for events such as supernovae that occur on day-or-longer timescales. However, these surveys are not sensitive to the very diverse class of shorter-timescale objects, including transiting exoplanets, young stellar variability, eclipsing binaries, microlensing planet events, gamma ray bursts, young supernovae, and other exotic transients, which are generally studied with individual small telescopes staring at single fields of view.

In this paper we explore an approach to reaching rare, short-timescale events across the sky: using a large array of telescopes to place a pixel on every part of the sky and integrating throughout the night to achieve depth. These systems have been prohibitively expensive up to now because of the extremely large number of pixels required to cover the sky with reasonable sampling, to say nothing of the logistics of building and maintaining the very large numbers of telescopes and storing the resulting data. The rise of consumer digital imaging and decreasing storage costs offer a solution to these problems. New surveys have exploited mass-produced compact CCD cameras and camera lenses to cover ever larger areas (e.g. \citealt{Vestrand2002, Pollacco2006, Pepper2007, Bakos2009, Malek2010, Law2013, Shappee2014}). The logical end-point of this approach is a survey which covers the entire sky with good pixel sampling; we here explore the science capabilities of such a system. 

\begin{figure}
  \centering
  \resizebox{1.0\columnwidth}{!}
   {
   \includegraphics[]{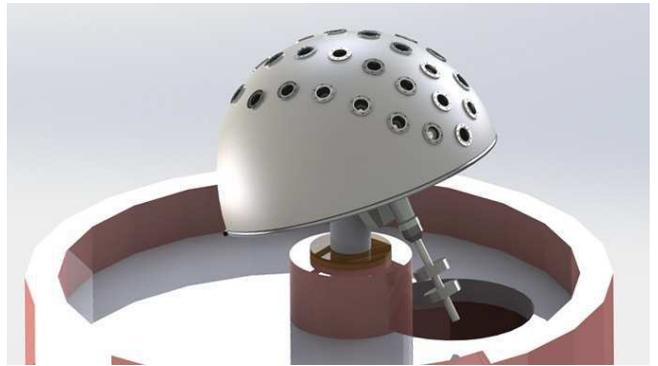}

   }
   \caption{The Evryscope all-sky gigapixel-scale telescope currently under construction at UNC Chapel Hill. The system consists of 27 individual telescopes based on commodity hardware, encapsulated in a 1.8m-diameter custom-molded dome which mimics the sky's hemisphere, and mounted on an off-the-shelf equatorial mount.}

   \label{fig:evryscope}
\end{figure}

To provide an example set of capabilities, we discuss the Evryscope (Figure \ref{fig:evryscope}), the array telescope we are currently constructing at the University of North Carolina at Chapel Hill. The system \citep{Law2012AAS, Law2012spie, Law2013, Law2014SPIEevryscope} is a low-cost 0.8 gigapixel robotic telescope that images 8,660 square degrees in each exposure. Contrasting the traditional telescope (Greek; far-seeing) to this instrument's emphasis on overwhelmingly wide fields, we have named our array telescope the Evryscope, from the Greek for wide-seeing.

The Evryscope is designed to open a new parameter space for optical astronomy, trading instantaneous depth and sky sampling for continuous coverage of much larger sky areas. An Evryscope is essentially a 10cm-scale telescope pointed at the entire accessible sky simultaneously. As such, these systems will complement large-telescope pointed surveys by enabling much shorter-cadence observations of much larger numbers of targets. Although we concentrate on the Evryscope array in this paper, other systems are under development, including Fly's Eye \citep{Vida2014} and HATPI (G. Bakos, private communication).

The ability to produce a realtime (few-minute-cadence) movie of the sky will enable realtime searches for transient and variable phenomena of all types, the photometric monitoring of millions of stars simultaneously, and can provide pre-imaging of unexpected events detected by other surveys. These capabilities have the potential of significantly contributing to many fields, a selection of which we describe in this paper. We summarize the science cases addressed in Table \ref{tab:science_cases}.

\begin{deluxetable}{ll}
\tablecaption{\label{tab:science_cases}All-sky gigapixel-scale telescope science cases}
\tabletypesize{\footnotesize}
\startdata
\bf{Science case} & \bf{Section}\\
\hline
\bf Bright, known objects & \S\ref{sec:bright}\\
\hspace{0.5cm}Transiting exoplanets & \S\ref{sec:transiting_exoplanets}\\
\hspace{1cm}Bright, nearby stars & \S\ref{sec:nearby_stars}\\
\hspace{1cm}Habitable planets around nearby M-dwarfs & \S\ref{sec:mdwarfs}\\
\hspace{1cm}White dwarf transits & \S\ref{sec:wd_transits}\\
\hspace{1cm}TESS planet yield enhancement  &  \S\ref{sec:TESS}\\
\hspace{0.5cm}Other exoplanet detection methods & \S\ref{sec:other_exoplanets}\\
\hspace{1cm}Transit and eclipse timing exoplanet detection & \S\ref{sec:eclipse_timing}\\
\hspace{1cm}Stellar pulsation timing exoplanet detection & \S\ref{sec:pulsations}\\
\hspace{1cm}Nearby-star microlensing & \S\ref{sec:microlensing}\\
\hspace{0.5cm}Stellar astrophysics & \S\ref{sec:stellar_var}\\
\hspace{1cm}Mass-radius relation &\S\ref{sec:mr_relation}\\
\hspace{1cm}Young stars & \S\ref{sec:young_stars}\\
\hspace{1cm}White-dwarf variability monitoring & \S\ref{wds}\\
\hspace{1cm}Variability from accreting compact objects & \S\ref{sec:accrete_var}\\
\hspace{1cm}Unexpected stellar events & \S\ref{sec:unexpected_stellar}\\
\bf Faint transient events & \S\ref{sec:extragalactic}\\
\hspace{0.5cm}Nearby, Young Supernovae &  \S\ref{sec:nearby_sne}\\
\hspace{0.5cm}Gamma-Ray Bursts & \S\ref{sec:GRBs}\\
\hspace{0.5cm}Gravitational wave EM counterparts & \S\ref{sec:ligo}\\
\hspace{0.5cm}Unknown or unexpected transients & \S\ref{sec:unknown_transients}\\
\end{deluxetable}

The paper is organized as follows. In \S\ref{sec:concept} we describe the astronomical regimes that currently-feasible all-sky telescopes can address; in \S\ref{sec:evryscope} we introduce the Evryscope as an example of such a system and detail the projected performance of our prototype system, as well as providing overviews of the data reduction design and the performance of our prototype systems which test the performance of individual Evryscope cameras. In the following sections we detail the science contributions such a system can make, to bright, known-object variability surveys (\S\ref{sec:bright}) and relatively faint transient surveys (\S\ref{sec:extragalactic}). We summarize in Section \ref{sec:summary}.

\section{Performance Metrics and Science Regimes}
\label{sec:concept}
The Evryscope concept multiplexes many small-aperture, low-cost telescopes to cover as much of the visible sky as possible. The telescopes are all mounted into a rigid hemisphere. The hemisphere acts as a proxy for the dome of the sky itself; when rotated at the sidereal rate, all the mounted cameras track the sky simultaneously. 

\subsection{Performance metrics}
With a fixed field of view on the order of the size of the entire accessible sky (8,000-12,000 square degrees depending on acceptable airmass), the remaining important variables for the instrument design are: 1) the aperture of the telescopes and 2) the number of pixels spread over the field of view.

A useful metric for the evaluation of all-sky surveys is how rapidly the sky can be covered to a particular SNR. This clearly scales linearly with the field of view of the survey, but evaluating the scaling of the SNR with the other quantities requires including the effects of an Evryscope-like design's relatively large pixel scale\footnote{set by the currently-feasible number of pixels to distribute across the sky; currently, roughly 0.5-3 gigapixels are feasible at the \$$10^6$ level with reasonable storage and computation requirements.}. We start with the standard imaging noise equation:

\begin{equation}
SNR = \frac{S_p T_{exp} A_{tel}}{ \sqrt{(S_p + S_s P^2) T_{exp} A_{tel}}}
\end{equation} 

where $S_p$ is the source flux in photons per second per square meter of collecting area; $T_{exp}$ is the exposure time; $A_{tel}$ is the telescope collecting area in square meters; $S_s$ is the sky background flux per square arcsecond; and P is the pixel scale in pixels per arcsecond. As we are aiming for a simple metric rather than a precise calculation, we simplify the expression by assuming that all sources are concentrated in single pixels and neglecting the effects of dark current, readout noise and quantum efficiency.

This relation can be simplified by exploring two regimes: bright targets where the sky background can be neglected, and faint targets where the sky background dominates the noise:

\begin{equation}
\rm{Bright\,source\,SNR} = \sqrt{S_p T_{exp} A_{tel}}
\end{equation} 

\begin{equation}
\rm{Background\,limited\,SNR} = \frac{S_p \sqrt{T_{exp} A_{tel}}}{ \sqrt{(S_s P^2)}} 
\end{equation} 

For a particular instantaneous field of view (FoV), in the bright source regime the time taken to cover the whole sky to a specified SNR and source brightness therefore scales as $\frac{1}{\rm{FoV} \times A_{tel}}$. In the faint-source regime the time taken scales with $\frac{P^2}{\rm{FoV} \times A_{tel}}$. 

\subsection{Science with all-sky telescope arrays}

From the above scaling relations we see that for sufficiently bright sources the pixel scale is not the limiting factor, and we recover the standard \'etendue metric: the telescope aperture times the field of view. For faint sources the pixel scale becomes the most important limiting factor, with the achievable SNR limited by the pixel scale squared.

These two performance regimes set where current relatively-large pixel all-sky telescope arrays can most effectively contribute: monitoring known bright objects where the pixel scale is relatively unimportant (once systematics are controlled for); and searching for rare short-variability-timescale objects, where the need to monitor large numbers of targets or large areas of sky rapidly is paramount. The source-magnitude crossover between the two regimes depends on the exact details of the hardware and site, but exploring the currently-feasible combinations for the systems described here we find the crossover typically occurs around $15^{th}$ magnitude.

\begin{figure}
  \centering
  \resizebox{1.05\columnwidth}{!}
   {
   \includegraphics[]{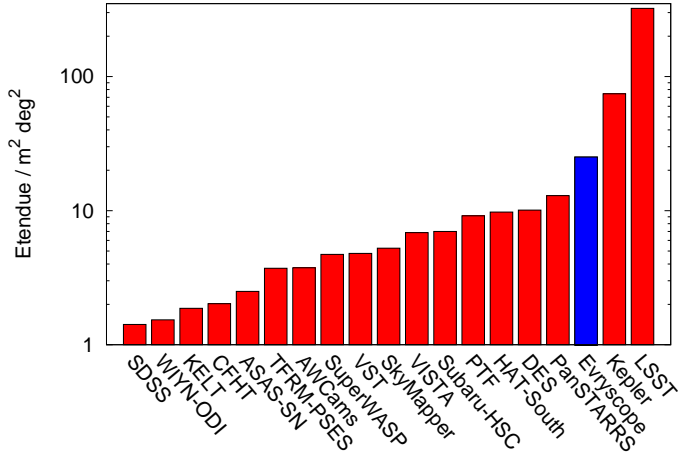}

   }
   \caption{The \'etendues of currently-operating general-purpose sky survey instruments (plus LSST; adapted from \citealt{Tyson2010}) compared to the Evryscope, the example all-sky array telescope described in this paper. We caution that the large pixel scale of the array telescope and other surveys based on similar technology makes this comparison only valid for bright sources which require high-cadence monitoring. To calculate the \'etendues for multiple-telescope surveys, we combine the telescope's FoVs where a survey has multiple sites and/or telescopes observing different fields; where a survey has multiple telescopes simultaneously observing the same field we combine their apertures.}

   \label{fig:etendues}
\end{figure}

In the bright source regime, the \'etendues of currently-feasible all-sky telescope systems exceed all current large-telescope sky surveys (Figure \ref{fig:etendues}; the enormous field of view effectively cancels their relatively small apertures). The arrays can therefore meet or exceed the performance of current sky surveys in the bright-source photometric-monitoring regime, including transiting exoplanets (\S\ref{sec:transiting_exoplanets}), microlensing events (\S\ref{sec:microlensing}), eclipsing binary and stellar pulsation monitoring for exoplanet detection (\S\ref{sec:eclipse_timing} and \S\ref{sec:pulsations}), and general stellar variability (\S\ref{sec:stellar_var}).

In the faint-source regime the sky background due to the large pixel scale becomes by far the limiting factor, suppressing the system performance by factors of tens or hundreds compared to the bright-star case; this means that all-sky telescope arrays are likely only to effectively contribute where the ability to achieve extremely high cadence over many objects (or sky areas) spread around the sky is of overriding importance. This includes extremely early-time observations of nearby supernovae (\S\ref{sec:nearby_sne}; optical observations of gamma-ray-bursts and orphan afterglows (\S\ref{sec:GRBs}); and cross-matching observations with high-cadence all-sky surveys operating at other wavelengths (\S\ref{sec:unknown_transients}) or spectra (\S\ref{sec:ligo}).

The cross-over between these two regimes is set by the pixel scale, and pushing to covering more pixels across the sky would push the crossover between "bright" and "faint" sources to much fainter levels. The pixel sampling is set by the number of detectors and telescopes that can be feasibly purchased and mounted, along with the data storage and reduction challenges. When the pixel scale pushes to similar levels to current sky surveys (around 1 arcsecond per pixel), an all-sky-telescope array with a similar \'etendue to those sky surveys becomes competitive at all magnitudes and timescales.  Moving to providing simultaneous arcsecond-scale sampling across the accessible sky would require 100s-of-gigapixels instruments, which will be challenging.

\section{The Evryscope}
\label{sec:evryscope}
As a concrete example of a currently-feasible all-sky telescope array, we here detail the Evryscope, currently under construction at UNC Chapel Hill. The Evryscope consists of a single hemisphere that contains twenty-seven 61mm-aperture telescopes, each with a rectangular 28.8MPix interline CCD imaging a 380 sq. deg. FoV using Rokinon 85mm F/1.4 lenses and including a five-element filter wheel. The hemisphere tracks the sky on a standard German Equatorial mount, imaging an instantaneous 10,200 sq. deg. FoV (8,660 sq. deg. when overlap between cameras is taken into account). The individual telescopes are fixed into holes in an aluminium-reinforced fibreglass dome. The interline CCDs provide an electronic shutter, so during normal operation the entire instrument operates with only one moving part: the RA drive. Each camera has a five-element filter wheel used as a dark shutter; the survey normally operates with a single filter, although we retain the option for future multi-filter surveys. All data is stored and analyzed on-site, with $\sim$100TB/year of compressed FITS images stored into network-accessible-storage (NAS) units. The hardware specifications are summarized in Table \ref{tab:evryscope}, including an estimate of the photometric performance of the system. We plan to deploy the Evryscope at the CTIO observatory in 2015.

\begin{deluxetable*}{ll}
\tablecaption{\label{tab:evryscope}The specifications of the Evryscope}
\tabletypesize{\footnotesize}
\startdata
\bf Hardware & \\
\hline
\noalign{\vskip 1mm}   
System design & 27 individual telescopes; shared equatorial mount\\
Telescope apertures & 61mm\\
Telescope mounting & Fibreglass dome w. aluminium supports\\
Detectors & 28.8MPix KAI29050 interline-transfer CCDs\\
& 7e- readout noise @ 4s readout time\\
& 50\% QE @ 500nm; 20,000 e- capacity w. anti-blooming\\
Field of view & 380 sq. deg. per telescope ($23.8^{\circ}\times16.0^{\circ}$)\\
              & 10,200 sq.deg. instantaneous total\\
              & 8,660 sq. deg. excluding overlap regions\\
Sky coverage per night & 18,400 sq. deg. (2-10 hours per night)\\
Total detector size & 780 MPix\\
Sampling & 13.3\arcsec/pixel\\
Observing strategy & Track for 2 hours; reset and repeat\\
Data storage & All data recorded for long-term analysis\\
             & 100TB / year (compressed)\\

\hline
\vspace{0.25cm}\\

\bf Performance &\\
\hline
\noalign{\vskip 1mm}   

PSF 50\% enclosed-energy diameter & 2 pixels in central 2/3 of FoV; 2-4 pixels in outer 1/3\\
Exposure times & 120s standard (plus shorter for bright-star mode)\\
Survey efficiency & 97\% efficiency from 4s camera readout\\
Limiting magnitude & $\rm{m_V}$=16.4 (3-sigma; 120s exposure)\\
& $\rm{m_V}$=18.2 (3-sigma; 1 hour)\\
& $\rm{m_V}$=19.0 (3-sigma, 1 night)\\
Photometric performance & 1\% photometry on $\rm{m_V<12}$ stars every 2 minutes (inc. scintillation)\\
                        & 3-millimag photometry on $\rm{m_V}$=11.5 stars every 20 minutes\\
                        & 3-millimag on $\rm{m_V}$=6 stars in 10 mins (saturation-limited short exps.)\\
                        & 1\% photometry on $\rm{m_V}$=15 stars every hour\\
\end{deluxetable*}

The photometric performance calculations in Table \ref{tab:evryscope} use conservative assumptions: median V-band sky brightness at a good dark-sky site of $\rm{m_V}$=21.8 (SDSS measurements; \citet{Abazajian2003}); atmosphere + telescope throughput 45\%; 50\% of encircled energy within a 4-pixel aperture, and a 25\% light-loss due to average vignetting and angular quantum efficiency effects across the field. 

The optimal survey filter depends on the exact science being targeted; for the purposes of this paper we assume a V-band filter will be used, although the filter wheels give flexibility during operations to perform multiple-filter surveys.

\begin{figure}
  \centering
  \resizebox{1.0\columnwidth}{!}
   {
   \includegraphics[]{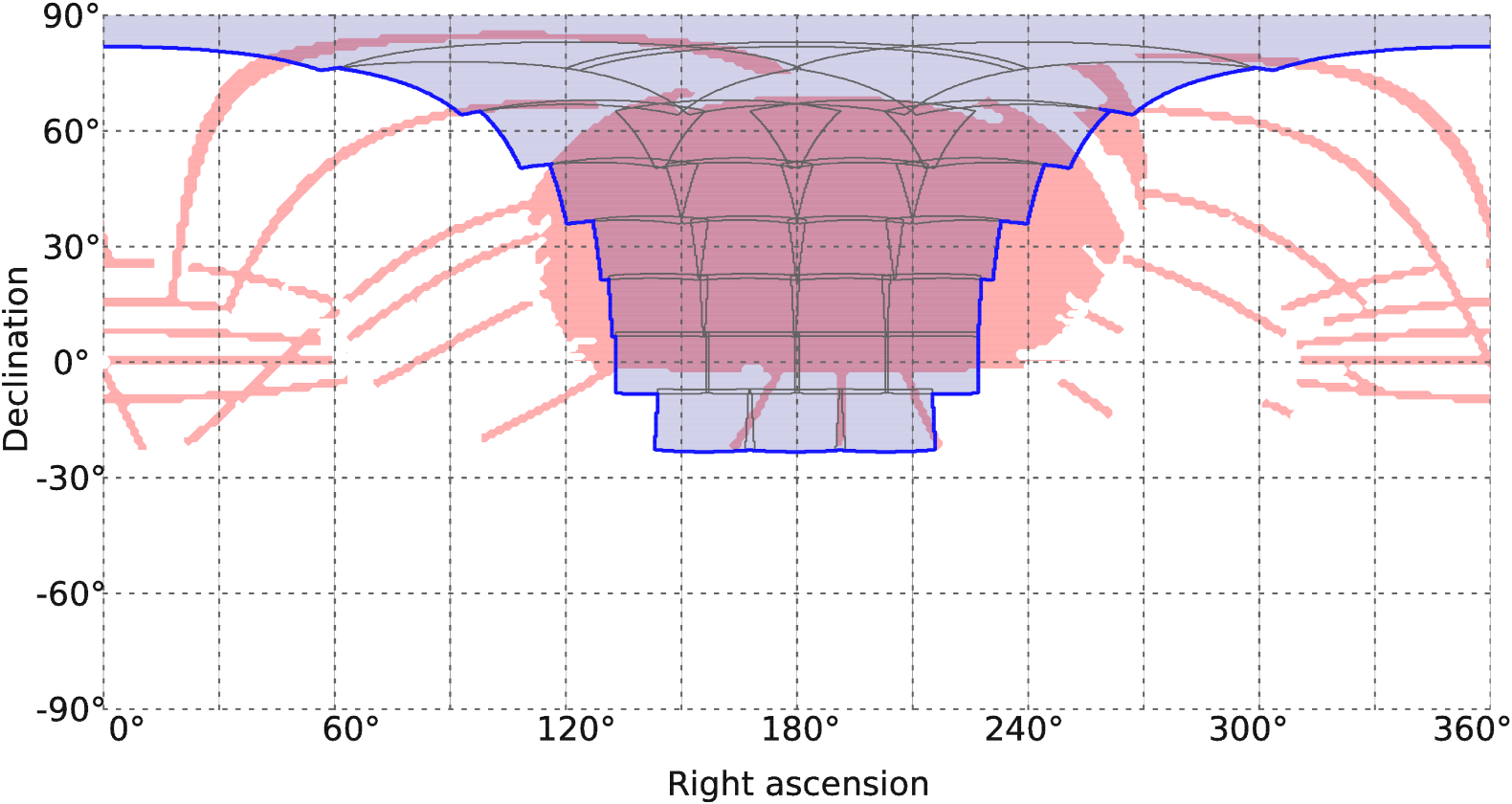}

   }
   \caption{The instantaneous Evryscope sky coverage, including the individual camera fields of view, for a mid-latitude Northern-hemisphere site ($30^\circ$N).  The SDSS DR7 photometric survey coverage \citep{Abazajian2009} is shown for scale. }

   \label{fig:sky_coverage}
\end{figure}

\subsection{Telescope array tracking modes}
\begin{figure}
  \centering
  \resizebox{1.0\columnwidth}{!}
   {
   \includegraphics[]{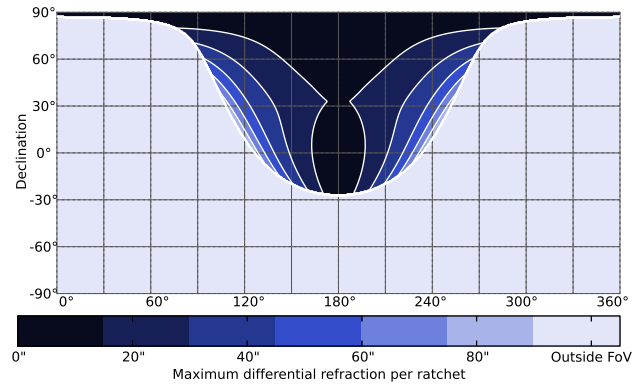}

   }
   \caption{The maximum atmospheric-refraction-induced change in stellar position over the course of a two-hour tracking ratchet, for a system with a field of view of $120^{\circ}$ located at a site with $33^{\circ}$N latitude. 28\% of the field maintains $<$13\arcsec motion; 56\% less than 26\arcsec (one PSF FWHM for the Evryscope system). We use the simple atmospheric model detailed in \cite{Meeus1991}, adjusted for typical observatory altitudes.}

   \label{fig:atmos_refrac}
\end{figure}

\begin{figure}
  \centering
  \resizebox{1.0\columnwidth}{!}
   {
   \includegraphics[]{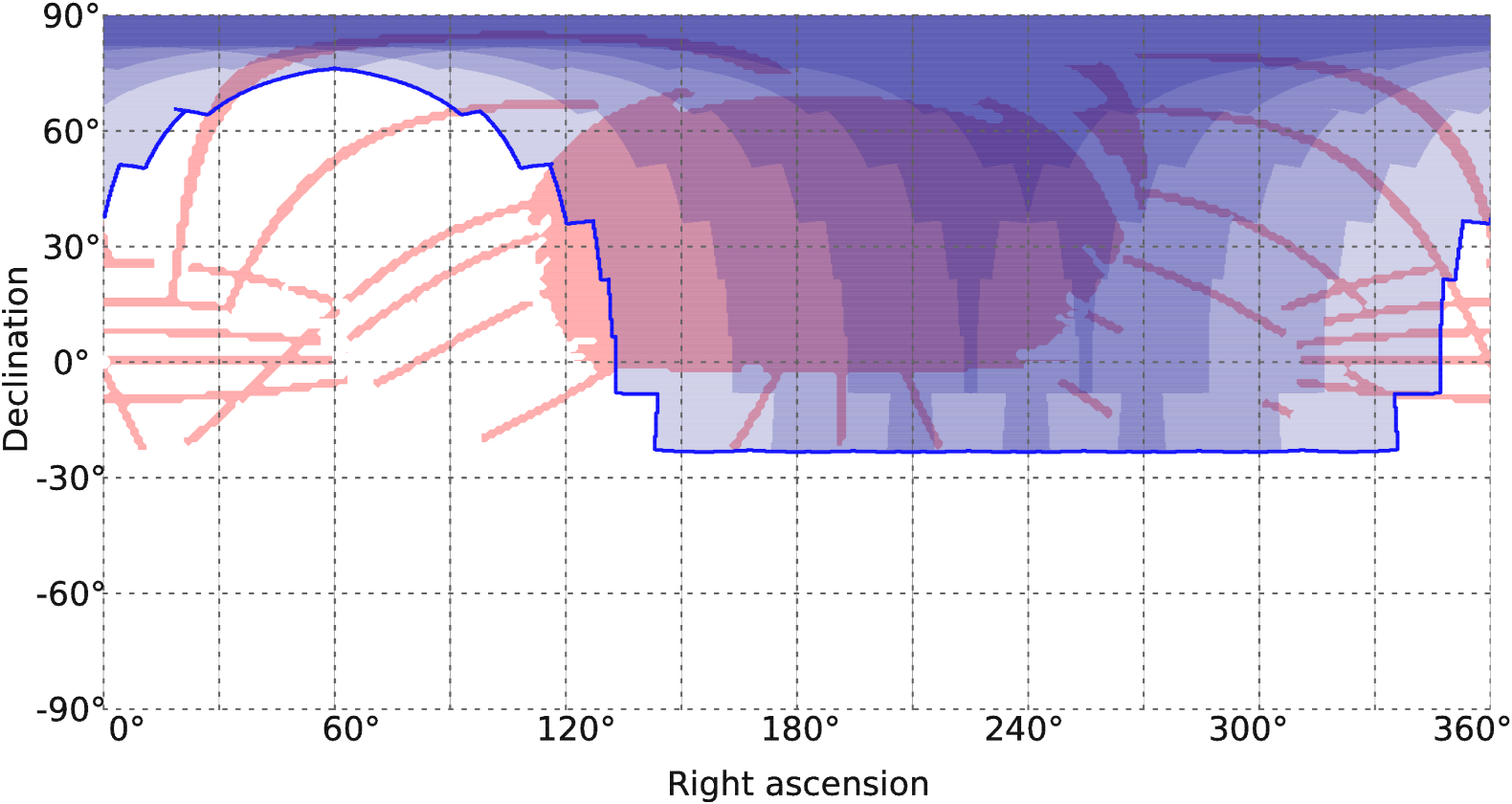}

   }
   \caption{The Evryscope sky coverage in one 10-hour night.  The intensity of the colouration corresponds to the length of continuous coverage (between 2 and 10 hours, in steps of two hours) provided by the ratcheting survey. For a mid-latitude Northern-hemisphere site ($30^\circ$N).}

   \label{fig:night_sky_coverage}
\end{figure}

The Evryscope will track 8,660 square degrees for two hours at a time before moving back (Figure \ref{fig:sky_coverage}); the field of view is wide enough that stars are continuously observed for an average of $\approx$6 hours each night (more at high declinations; Figure \ref{fig:night_sky_coverage}). In total 18,400 square degrees are continuously observed for at least two hours each night. This ``ratcheting'' survey setup is designed for both exoplanet transit searches (precision long-term photometry) and co-adding of images for deep imaging (transient searches). To aid in both precision photometry and co-adding, the ratcheting survey maintains PSF-width-level positioning for $\approx$56\% of the FoV on two-hour timescales (limited by atmospheric refraction in part of the field; Figure \ref{fig:atmos_refrac}).

An alternate approach would be to track the system for the length of an exposure only and ratchet back on each exposure. However, this approach leads to stars moving across the telescopes' FoVs from one exposure to the next, leading to potentially large PSF changes with the current commercial camera lens optical qualities, and thus has the potential to greatly increase systematic errors in the photometry. The PSF differences are up to factors-of-two in FWHM from the center to the edges of each telescope's FoV, and vignetting can also vary by up to  50\% \citep{Law2013}. In a few-minute-tracking mode, this change is large enough to produce rapid changes in the SNR of the stars and the limiting magnitude of the system on each part of the sky, also leading to potential data reduction and survey performance issues. 

For these reasons we have designed the Evryscope for two-hour tracking, simplifying the photometric reduction and keeping the PSFs and limiting magnitudes stable for at least 60 measurement epochs in each tracking period. The two-hour-tracking mode also guarantees that each part of the sky is covered for at least two hours each night, simplifying time-series photometry. 

There are two possible tracking surveys: 1) observing new field centers each night, slightly shifted in RA; and 2) maintaining field centers for longer periods at the cost of observing at less-optimal airmasses. The first introduces extra systematic noise from the night-to-night changes of the delivered PSF and vignetting on each star; the second introduces extra airmass-produced systematics. We will decide the optimal strategy on the basis of on-sky systematics measurements.

\subsection{Evryscope data reduction}
The Evryscope data will be stored and analyzed on-site. The system will generate approximately 100TB/year of compressed FITS images stored into low-cost network-accessible-storage (NAS) units. The on-site analysis pipeline, based on the pipelines developed for the Evryscope arctic prototype cameras (\S\ref{sec:arctic}), astrometrically and photometrically calibrates images, extracts sources and then associates them with a reference catalog made from previous Evryscope epochs. The current version of the pipeline can keep up with the incoming Evryscope data stream using a 12-core server, allowing realtime data analysis. Imaging data will be stored on-site and then physically transferred on hard disk drives for further analysis every few months. On shorter timescales, the source-associated light-curves will be transferred by Internet to a cluster at UNC Chapel Hill for detrending and transit detection.

The realtime source association tests allow quick detection of obvious new objects which may require rapid follow-up, such as supernovae (\S\ref{sec:nearby_sne}) and GRBs (\S\ref{sec:GRBs}). The pipeline stores the data in a compressed format which allows rapid retrieval of time-series cut-out images of interesting areas of the sky, also allowing easy after-the-fact measurements of transients detected by other surveys. Later developments of the pipeline will include realtime differential image analysis to detect new sources in crowded regions of the sky. The detailed design of the Evryscope pipeline will be described in a future publication.

\subsection{Evryscope Performance Testing}
\subsubsection{AWCams: High Canadian Arctic Planet-Search Telescopes}
\label{sec:arctic}
The AWCams (Arctic Wide-field Cameras) are two small telescopes designed to search for exoplanet transits around bright stars (V=5-10). They are similar to individual Evryscope cameras except that they lack the tracking that enables long exposures.  The cameras are deployed at the PEARL atmospheric science laboratory at 80$^{\circ}$N in the Canadian High Arctic, where continuous winter darkness greatly increases their sensitivity to long-period exoplanets.  The cameras, the AWCam project, and the attained performance are described in detail in \cite{Law2012spie, Law2013, Law2014_arctic}.

The AWCams have been operating in the Arctic for three winters, including a test run in February 2012 and full-winter operations in the 2012/13 and 2013/14 winters. The robustness of the hardware and enclosure design has been validated by essentially uninterrupted operation throughout the entire deployment period, including a total of 10 months of completely unattended robotic operation. Throughout the winters the cameras kept themselves (and crucially their windows) clear of snow and ice, took over 60TB of images, and consistently maintained few-millimagnitude photometric precisions over several-month timescales \citep{Law2014_arctic}. This performance with similar lenses and CCDs to Evryscope hardware demonstrates that exoplanet-detection-level photometric precisions are achievable with our software pipelines even in relatively-hard-to-reduce untracked data.

\subsubsection{Tracking Single-Telescope Prototype}

We have also built and operated a prototype tracking unit-telescope system based at the Appalachian State University Dark Sky Observatory (DSO). The individual Evryscope telescope was mounted to a Celestron CPC1100 using a custom-built dovetail plate and wedge, providing tracked exposures over the two-minute Evryscope timescales with a tolerance small enough ($<$ 1 pixel) to emulate the final Evryscope performance. The dark-sky conditions at DSO have allowed us to verify that the limiting magnitude and image quality calculations described above match the actual performance of the system under observatory conditions. 

\subsection{Summary of Evryscope Performance}
In the targeted range of declinations (110-degrees of declination) the Evryscope will generate a dataset including:

\begin{enumerate}
\item{two-minute-cadence multi-year light curves for every star brighter than $\rm{m_V}$=16 in the target range of declinations.}
\item{millimagnitude minute-cadence photometry for every star brighter than $\rm{m_V}$=12 in the target range of declinations.}
\item{minute-by-minute record of all events in the sky down to $\rm{m_V}$=16.5, with only 3\% deadtime for image readout.}
\item{$\rm{m_V}$=19 in one-hour integrations; every part of the sky observed for at least 6.5 hours per night.}
\end{enumerate}

\begin{figure}
  \centering
  \resizebox{1.0\columnwidth}{!}
   {
   \includegraphics[]{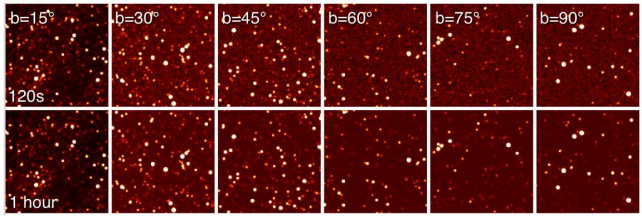}

   }
   \caption{We evaluated the crowding levels for the Evryscope by developing a tool to simulate images on the basis of USNO-B1 photometry (Monet et al. 2003), including the effects of crowding, the lens PSFs, the detector sampling, and photon and detector noise. We show above representative 10-arcminute fields in single exposures and in one-hour co-adds, scaled to show the faintest detectable stars. Crowding does not limit photometry of at least 90\% of stars above 15 degrees galactic latitude in 120s exposures (30 degrees in 60m exposures).}

   \label{fig:crowding}
\end{figure}

The Evryscope's 13''/pixel image scale is small enough to allow separated precision photometry for at least 90\% of stars observed above 15 degrees galactic latitude (Figure \ref{fig:crowding}). 

This large dataset will allow a wide range of science to be performed simultaneously; we detail potential scientific contributions for Evryscope-like systems in the following sections. To simplify the descriptions of the accessible targets, we  assume that multiple systems will be deployed to cover both the Northern and Southern hemispheres, and so refer to covering all stars of particular types, rather than limiting to a particular hemisphere. We provide a quick-reference to the science cases in Table \ref{tab:science_cases}.

\section{Bright, known-object regime: exoplanets and stellar variability}
\label{sec:bright}

The Evryscope's unique contribution to exoplanets will be its ability to monitor extremely large sky areas at high-cadence. It will thus simultaneously cover large numbers of rare, widely-separated targets that would otherwise require individual telescopes for each target. This will enable large surveys for rare transiting objects as well as providing the datasets for other more exotic exoplanet-detection methods.

\subsection{Transiting Exoplanets}
\label{sec:transiting_exoplanets}
The Evryscope's uniquely large field of view enables transit searches in stellar populations that, because of their rarity, have been inaccessible up to now. Compared to current transiting planet surveys (Table \ref{tab:current_transits}), the Evryscope has at least 15 times the instantaneous field of view, at least five times the number of pixels, and similar sky sampling.   Crucially, the photometric performance on bright stars is scintillation-limited, and so the ultimate photometric performance of the system is set by the aperture and the length of time that each star is covered during transit times. The Evryscope's continuous coverage enables averaging over more than 100 datapoints and more than an hour for each transit occurrence, enabling much improved photometric performance compared to surveys which cover large sky areas by imaging smaller areas in sequence.

\begin{deluxetable*}{llllllll}
\tablecaption{\label{tab:current_transits}The Evryscope compared to current transiting planet surveys}
\tabletypesize{\footnotesize}
\startdata

\bf Survey & \bf FoV\tablenote{Maximum simultaneous FoV over all sites} & \bf Aperture & \bf \arcsec/pixel & \bf Sites & \bf MPix/site& \bf Targets & \bf Ref.\\
\bf  & \bf / sq deg. & \bf / mm & \bf  & \bf & \bf & \bf & \bf \\
KELT & 676 & 42 & 23.0 & 2 & 16  & Bright stars & \citet{Pepper2007}\tablenote{Multiple sites observing different fields}\\
SuperWASP & 488 & 111 & 13.7 & 1 (8 tels.) & 32  & General transits & \citet{Pollacco2006}\\
HAT-South & 128 & 180 & 3.7 & 3 (6 tels.) & 128 & General transits & \citet{Bakos2009,Bakos2012}\tablenote{Multiple sites with continuous observation of 128 sq. deg.}\\
MEarth & 2.8 & 400 & 0.8 & 2 (8 tels each.) & 32  & M-dwarfs & \citet{Irwin2009}\tablenote{Multiple sites observing different fields}\\
CSTAR & 20 & 145 & 15 & 1 (Antarctic) & 4  & Bright stars & \citet{Wang2014}\tablenote{Antarctic site with continuous observations during winter}\\
TFRM-PSES & 19 & 500 & 3.9 & 1 & 16 & M-dwarfs & \citet{Fors2013}\\
AWCams & 504 \& 1295 & 71 \& 42 & 22 \& 36 & 1 (Polar) & 32 & Bright stars & \citet{Law2013}\tablenote{Arctic site with continuous observations during winter}\\
NGTS & 96 & 200 & 1.1 & 1 (12 tels) & 48 & Bright stars & \citet{Wheatley2013}\\
Kepler & 105 & 950 & 4.0 & 1 & 95  & Earth-like planets & \citet{Koch2010}\tablenote{Space-based telescope with continuous observations}\\
Evryscope & 8,660 & 61 & 13.3 & 1 & 780  & Bright stars, MDs, WDs & \citet{Law2012spie}\\

\end{deluxetable*}

\begin{figure}
  \centering
  \resizebox{0.9\columnwidth}{!}
   {
   \includegraphics[]{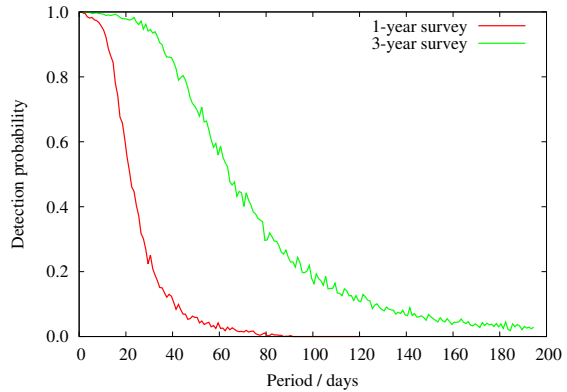}

   }
   \caption{The probability of detecting at least three significant eclipse/transit events in an Evryscope survey, for 1\%-level transits and eclipsing binaries with highly significant detections in each datapoint. We assume 33\% weather losses and simulate the observing window effects for a 20-degree-declination target observed from our planned Evryscope site. The detection efficiency for longer-period, low-transit-depth planets will be reduced by the number of data points to be searched.}

   \label{fig:planet_detect}
\end{figure}

\subsubsection{Bright, nearby stars}
\label{sec:nearby_stars}
Follow-up observations of transiting exoplanets, by either emission spectra during secondary eclipse \citep{Madhusudhan2011, Knutson2007} or transmission spectroscopy \citep{Sing2011,Snellen2010,Tinetti2007,VidalMadjar2003,Charbonneau2002} techniques, have revealed direct measurements of albedos, atmospheric composition, chemistry, and even phase curves showing features on the planetary cloud layers. These observations have been performed for only a very few planets, however, because they require one thing above all else: photons from the host star. Without a star significantly brighter than most narrow- field transit searches can currently monitor, it is prohibitively expensive to reach sufficient signal-to-noise on the most interesting spectral features of the planet transit even with future observatories such as JWST \citep{Madhusudhan2014, Rauer2011, Kaltenegger2009}. Current exoplanet transit surveys (Table \ref{tab:current_transits}) are limited to fields of view in the few-hundred-square-degree range at best and cannot effectively search for transits around a large sample of rare bright stars.

The Evryscope will have an order-of-magnitude larger field of view than the next-largest current exoplanet surveys, allowing a long-term transit survey covering 70,000 stars brighter than $\rm{m_V}$=9 \citep{Perryman1997} (the brightest stars will be observed in a short-exposure supplement to the standard Evryscope survey strategy). Given the Kepler-measured planetary population rate (e.g. \citealt{Howard2012}) and detailed observing efficiency, weather window, geometric and false-positive corrections (Figure \ref{fig:planet_detect}; \citealt{Law2013}), we estimate that the Evryscope will at least double the known number of transiting giant planets around stars brighter than $\rm{m_V}$=9.

\subsubsection{Rocky planets in the habitable zones of M-dwarfs}
\label{sec:mdwarfs}
Despite being the most common stellar type in our galaxy, the transiting planetary population around M-dwarfs has not yet been explored in detail because of their extreme faintness compared to solar-type stars. Kepler can only cover a few faint thousand targets in this mass range (e.g. \citealt{Muirhead2012, Dressing2013}) and individually-targeted surveys like MEarth are also limited to a few thousand bright M-dwarfs at most \citep{Berta2013}. The results of a large exoplanet transit survey covering 1,000,000 M-dwarfs \citep{Law2012} and radial velocity surveys of a few hundred of the brightest M-dwarfs (e.g. \citealt{Butler2006, Endl2006,  Bonfils2013}) suggest that giant planets are extremely rare around M-dwarfs, although rocky planets seem to be much more common (e.g. \citealt{Howard2012}). The one transiting planet detected around a relatively bright M-dwarf, GJ1214b \citep{Charbonneau2009}, has been a subject of huge interest, with dozens of characterization papers published. 

The next step is to push towards the detection of large samples of the apparently common rocky planets around M-dwarfs, comparing their population statistics and ultimately composition and mass-radius relation to those of planets around solar type stars. Recent Kepler planetary population statistics suggest that the nearest transiting rocky planet in the habitable zone of an M-dwarf is less than 9pc away from us \citep{Dressing2013}. However, to have a chance of finding these planets around M-dwarfs bright enough and nearby enough to use for characterization, a large sample of nearby, bright M-dwarfs must be covered. In turn, their random and sparse distribution across the sky means we must cover a very large sky area to reach a significant number of targets. The Evryscope will be capable of simultaneously monitoring all bright, nearby late K stars and M-dwarfs for transiting rocky planets. A number of recently released all-sky nearby M-dwarf catalogs provide the starting target lists for this bright M-dwarf survey: 2,970 for CONCH-SHELL \citep{Gaidos2014}, 8,479 in \citep{Frith2013} and 8,889 in \citep{Lepine2011}.

With few-millimagnitude photometric precision, the Evryscope will be sensitive to planets as small as a few Earth radii around the small-radius mid-M-dwarfs. On the basis of Kepler planet statistics we estimate that a long-term Evryscope survey would detect 5-10 small planets around bright M-dwarfs – with good sensitivity to targets in the relatively small-radius habitable zones of those faint stars. In addition, the system will simultaneously search for giant planets transiting a 100-times larger sample of late M-dwarfs. This capability will be highly complementary to the new generation of infrared radial velocity planet surveys that will come online at roughly the same time as the Evryscope survey (e.g. \citealt{Quirrenbach2010,Artigau2014}).

\subsubsection{White dwarf transits}
\label{sec:wd_transits}
White dwarfs are an attractive exoplanet transit-search target because their small size enables the detection of extremely small objects -- rocky planets can occult the star, moon-sized objects give 10\%-range transit signals, and large asteroids may be detectable \citep{Drake2010, Agol2011}. The detection of a transiting planet around a white dwarf would give intriguing insights into the characteristics of planets in such an extreme environment; the planetary system evolution during the star's red giant phase; and possibly constraints on the properties of very small rocky bodies. However, the small white dwarf radius greatly reduces the geometric transit probability, even for hypothetical close-in planets; the small size also reduces the transit time to minutes. Because white dwarfs are very faint it has been very difficult to obtain a reasonably large sample of white dwarfs in a transit survey at all, let alone at the required minute cadences. Furthermore, the short transit lengths require a very high observational duty cycle on each target to have a reasonable chance of detecting the transits.

The Evryscope's all-sky all-the-time survey will be the first to be able to cover a very large sample of relatively bright white-dwarfs; because the white-dwarfs are so small, even faint targets can be effectively searched for rocky-planet transits by searching for drop-outs where the white dwarf disappears for a few minutes. Even finding one or two highly-significant dips would make a white dwarf transit candidate worth following up with other facilities. 

From current catalogs of bright white dwarfs (see \S\ref{wds}) we estimate that the Evryscope will be able to simultaneously cover hundreds of white dwarfs with better than 10\% photometric precision in each two-minute exposure, and $\sim10^3$ white dwarfs each night. This will enable us to place limits on the populations of -- or even discover -- Mercury-sized objects. As a by-product of this exoplanet transit search around white dwarfs, the Evryscope will also detect new eclipsing white dwarf binaries and periodically variable white dwarfs (\S\ref{wds}).

\subsubsection{Planet yield enhancement for TESS and other exoplanet surveys}
\label{sec:TESS}
The \textit{TESS} mission, the follow-on to the \Kepler mission, will cover the entire sky, searching for rocky transiting exoplanets around more 200,000 bright, nearby stars. With its four cameras with a $24{\degr} {\times} 24{\degr}$ field of view each, the mission requires a median stare-time on each part of the sky of months. With multiple-year surveys covering a large fraction of the entire sky simultaneously, the Evryscope will offer a highly complementary dataset to TESS.

\subsubsection{TESS host characterization photometry} TESS is currently planned for launch in 2017 \citep{Ricker2014}; by the time TESS is operational the Evryscope will have collected a three-year dataset on every  star in the TESS survey. The three-year Evryscope dataset will enable a sensitive high-cadence search of every TESS target for eclipsing binaries (of all periods), flare stars, exotic binaries, rotational modulation, sunspots, and all other intrinsic photometric variability, on timescales similar to the TESS cadence.

\subsubsection{Increasing the TESS planet yield}
By the TESS launch the Evryscope's dataset will contain at least one transit event from almost every giant planet transiting TESS targets, in orbital periods up to several months (and given the number of stars surveyed, a large number of well-sampled single-transit events from much-longer-period planets). This offers the opportunity of using the Evryscope dataset to confirm long-period planets that may only transit once during the TESS 60-day stare period. The Evryscope could thus greatly improve the TESS long-period planet yield, especially because we can use the TESS single-transit data to pick out shorter lengths of Evryscope data to search for transits, decreasing the required significance of individual detections in Evryscope data.

\subsubsection{Gaia}

Gaia typically revisits each field 70 times over its 5 year mission.  Given this revisit configuration and other constraints, Gaia is expected to astrometrically detect $\sim$2,000 Jupiter-size planets within than 200pc, most orbiting around bright GK dwarfs stars with periods ranging between 1.5 to 9 years \citep{Sozzetti2011, deBruijne2012}. In addition, it has been suggested that the photometry obtained during these revisits could allow the detection of a comparable number of planets in single-datapoint-per-transit detections \citep{Voss2013}. Although the number of targets covered is large, the false-positive concentration in this low-cadence data will make confirmation difficult and the low cadence will produce a low probability of detection for even short-period exoplanets. The Evryscope's photometric precision is comparable to Gaia's photometric precision (2mmag at $\rm{m_G}=12$; \citealt{deBruijne2012}), although the Evryscope's cadence is almost 20,000 times faster. The Evryscope's higher cadence will allow it to achieve higher detection sensitivities for short period planets and confirm long-period planets detected in Gaia multi-year dataset.

\subsection{Other exoplanet detection methods}
\label{sec:other_exoplanets}
\subsubsection{Transit and eclipse timing for exoplanet detection}
\label{sec:eclipse_timing}
Transit timing variations allow us to use changes in eclipse times to measure the influence of other bodies in a system on the transiting/eclipsing body's orbit. Measurements of the variations have been successfully applied to the confirmation of multiple-planet systems detected by Kepler (e.g. \citealt{Mazeh2013}). Similar timing variations have recently produced possible detections of planets orbiting eclipsing binary stars (e.g. \citealt{Potter2011, Marsh2013}), including faint stellar types such as white dwarfs that are not easily amenable to other planet survey methods such as radial velocities and reflected-light direct imaging.  An all-sky telescope will record minute-cadence lightcurves for every eclipsing binary brighter than $\rm{m_V}$=16.5. With multiple-year, every-night coverage when the targets are up, we will automatically obtain hundreds of precise eclipse-times for the thousands of short-period objects in the FoV. Compared to the current standard approach of selecting and monitoring individual interesting targets on longer timescales \citep{Marsh2013} this massively-multiplexed eclipse-timing survey will enable a much larger and more comprehensive eclipse-timing search for exotic planetary systems.

\subsubsection{Stellar pulsation timing for exoplanet detection}
\label{sec:pulsations}
Stellar pulsations can also serve as accurate clocks for discovering planets (see \citealt{sch10}). Confirmation of the pulse timing method's ability to find unseen companions has been provided in several cases (e.g., \citealt{vin93,bar11}), although none of the detected objects were planets. A few substellar and planetary detections have been reported, but without radial velocity confirmation \citep{sil07,mul08}.  The pulse timing technique is most sensitive to planets when (i) the pulsations have relatively short periods, from minutes to several hours; (ii) multiple high--amplitude, independent modes are present and well--separated in frequency space; and (iii) the pulsation periods are adequately stable, preferably to 1 part in 10$^{8}$ or better.   Objects with pulsation characteristics best meeting these criteria include the hot subdwarfs, white dwarfs, $\delta$ Scutis, and roAp stars, for example. The Evryscope's cadence is well--suited to monitoring these type of pulsations, although for the shortest-period hot subdwarf and white dwarf pulsators, they might appear in the super-Nyquist regime.   While other planet detection methods quickly lose their utility at larger separation distances and longer orbital periods, the pulse timing method remains relatively robust in this regime, as it depends primarily on the host star's overall {\em displacement} from the barycentre (and not a perfectly edge--on orbital alignment or large radial velocity). The Evryscope should provide pulse timings for thousands of pulsators that are sensitive to planetary--sized objects.

\subsubsection{Nearby-star microlensing}
\label{sec:microlensing}
Typical galactic microlensing events occur on week-timescales, but exoplanets orbiting the lens star (or even isolated planets) can be detected as much shorter timescale bumps in the light curves. Most microlensing surveys (for example, OGLE; \citealt{Udalski2008}) have been performed with larger telescopes observing relatively small fields towards the galactic plane, where there is a large population of background stars for lensing. However, occasional spectacular events around relatively nearby stars (e.g. \citealt{Gaudi2008}) have demonstrated that a sufficiently large-area survey has the opportunity to detect much closer events -- and detect planets smaller than Earth in half-AU orbits \citep{Gaudi2008}.  The key to successful microlensing planet detection is continuous monitoring and rapid follow-up. The Evryscope's few-minute temporal resolution, high photometric precision and all-sky coverage mean that planetary signatures will be directly visible in the light curves (this has recently been demonstrated in smaller fields by \citealt{Shvartzvald2013}). A survey with the Evryscope's sky coverage is expected to detect several near-field microlensing events each year, along with many more conventional distant events towards the galactic plane \citep{Gaudi2008, Han2008}.

\subsection{Stellar Astrophysics}
\label{sec:stellar_var}
With two-minute-cadence monitoring of every star brighter than $\sim\rm{m_V}$=16.5, all-sky array telescopes will enable the discovery and characterization of a wide range of stellar variability.

\subsubsection{Mass-radius relation from eclipsing binaries}
\label{sec:mr_relation}
The Evryscope will provide a complete full-sky inventory of eclipsing binary systems with orbital periods of $P \la 60$ days, greatly  expanding the number of systems which are amenable to measurements of stellar radii and dynamical masses. These measurements are crucial for the study of the stellar mass-radius relation, which is currently uncertain at the 10\% level for K-M stars (e.g. \citealt{LopezMorales2007, Boyajian2012}) and directly carries through to uncertainties in models of stellar evolution \citep{Chabrier2007, Morales2010, Feiden2013} and determinations of the radii of transiting extrasolar planets \citep{Fortney2007, Charbonneau2007, Swift2012}. Previous results have shown that stellar radii could be biased by stellar activity \citep{LopezMorales2007} and rotation \citep{Kraus2011}, which argues that long-period eclipsing binary systems ($\gsim$10 days period, which are not tidally locked and can rotate at the same velocity as single field stars) will be crucial for determining the true mass-radius relation. A small set of long-period K-M systems were identified by Kepler \citep{Prsa2011}, thanks to its 100\% duty cycle, but it could only survey a small area of the sky. Long-period systems have otherwise been largely neglected by previous variability surveys, which do not have a significantly high long-term duty cycle to identify the occasional eclipses of long-period systems. The Evryscope will achieve a more complete inventory of such systems over the entire sky, making an unprecedented contribution to the mass-radius relationship for very low mass stars (e.g. \citealt{Law2012, Zhou2014}).

\subsubsection{Young stars}
\label{sec:young_stars}
Stellar variability is ubiquitous among young stars (e.g. \citealt{Skrutskie1996, Carpenter2002}). Newly-formed stars were first identified as a class from their variability, a feature which is still recognized in their name ({\em T Tauri} stars; \citealt{Joy1945}; \citealt{Herbig1962}). This variability is driven by stochastic brightness variations from the accretion of circumstellar material (as for T Tauri itself) as well as quasi-periodic rotational modulation from spots (as in BY Draconis stars). This variability was crucial in compiling early catalogs of young stars (e.g., \citealt{Kenyon1995}), but over the past 15 years, it has been largely supplanted by wide-field space-based surveys in the mid-infrared (e.g., \citealt{Evans2009}). However, these surveys are only sensitive to stars which host protoplanetary disks or envelopes; the disk-free population has remained largely unidentified. Even youth indicators like X-Ray and UV emission (e.g., \citealt{Wichmann1996, Scelsi2007, Findeisen2010, Shkolnik2011}) remain swamped by contamination from field intermediate-age stars, spectroscopic binaries, and chance alignments with background extragalactic sources. The Evryscope will directly identify disk-free young stars based on their spot-driven variability (e.g. \citealt{Cody2014}), which can achieve photometric amplitudes of $\sigma \sim 0.1$ mag in the optical for stars younger than 100 Myr \citep{Herbst2002, Cody2013}.

\subsubsection{White-dwarf variability monitoring}
\label{wds}
The per-exposure detection limit of our program is roughly equal to that of the Edinburgh-Cape (EC) Blue Object Survey \citep{stobie97}.  Zones 1 and 2 of the EC survey cover 3,290 sq deg of the southern sky and include 229 white dwarf stars \citep{kilkenny97, odonoghue13}.  Scaling from this surface density, we expect to monitor more than 600 white dwarfs with every exposure, and more than 1,000 each night.  These data will be sensitive to pulsations, rotation, and various binary phenomena.
 
In certain ranges of temperature, white dwarf stars experience non-radial g-mode pulsations that result in photometric variations having periods between $\sim$$1.5$ min and 30 min with amplitudes ranging from 0.1\% to 10\% \citep{winget08, fontaine08, althaus10}. Though this variation will not be directly visible in the Evryscope light curves of most of these stars, for many of them, it will announce itself by excess scatter in the light curves and will provide candidates for follow up time-series photometry and asteroseismic analysis.  Among the hydrogen-atmosphere pulsators (the ZZ Cetis) the 2-min integration times will mean Evryscope is relatively insensitive to the hotter pulsators, which tend to have periods from 100-200 s and amplitudes $\sim$1\%, but we will generally be able to detect the cooler pulsators because of their longer periods and larger amplitudes.  At $\rm{m_V=15.5}$, a typical 600 s oscillation with an amplitude of 1.5\% will be detected in one season of observing, while at the single-exposure detection limit, signals at the same period greater than 3.5\% will be detected.  We note as an example that in one season of observing, Evryscope data will constrain the phase of the dominant mode of the cool ZZ Ceti BPM 31594 ($\rm{m_V=15}$) to $\lesssim$ 3 s, sufficient, over time, to place constraints on cooling rates and orbiting planets \citep{kepler05, mullally08}.
 
The Evryscope will also be sensitive to rotation in some white dwarf stars.  Most white dwarfs presumably rotate, but it is often difficult to detect the rate of rotation, which is important for understanding angular momentum loss on the AGB.  White dwarfs with magnetic fields can show spots on their surfaces that result in photometric variability as they rotate \citep{brinkworth13}.  The detected periods range from 725 s \citep{barstow95} to years.  These stars will not only be useful as probes of rotation but will also be candidates for spectroscopic and polarimetric follow up to confirm and study their magnetic fields, and stable rotators can be used as probes of motion resulting from an orbiting companion \citep{lawrie13}.
 
White dwarfs with binary companions produce various types of photometric oscillation.  The secondary can show periodic variation resulting from reflection effect and ellipsoidal variations, and, of course, the system may be eclipsing, which can, among other things, provide important constraints on the white dwarf mass-radius relationship.  For some binaries, such as the bright ($\rm{m_V=12}$) white dwarf + hot subdwarf CD$-$30$^\circ$11223 \citep{geier13}, Evryscope data should determine the phase of the ellipsoidal modulation to $\sim$3.5 s every observing season.  The change of period of this system resulting from gravitational wave radiation is $6\times10^{-13}\,{\rm s\,s^{-1}}$.  Given the above phase precision, Evryscope data could detect this change in approximately a decade.

\subsection{Variability from accreting compact objects}
\label{sec:accrete_var}
In recent years it has started to become clear that accretion onto
compact objects is a relatively universal process, with global
similarities in the accretion process in disks around supermassive
black holes, stellar mass black holes, neutron stars, and white dwarfs
(see e.g. \citealt{McHardy2006, Klis1994, Scaringi2014}).  The new combination of high time resolution and high duty
cycle of observation has opened new parameter
space for studies of cataclysmic variables, in particular.  The Evryscope will offer three major advantages over Kepler -- the
ability to observe stars which are not, {\it a priori}, recognized as
interesting; coverage of a much wider part of the sky (which is
important for observing rare objects); and potentially longer time
baselines.  Like LSST, it will be able to detect outbursts from cataclysmic variables and X-ray binaries, but it may also be able to detect a possible hidden population of outbursts from short orbital systems, which should have especially short outbursts \citep{Knevitt2014}.

\subsubsection{Aperiodic variability from accretion flows}

The results on cataclysmic variables from Kepler highlight what can be
done with the Evryscope.  \citet{Scaringi2014} has shown, for example, that with
long, well-sampled Kepler observations, it is possible to make studies
of cataclysmic variables' power spectra that can be compared very well
to those made for X-ray binaries.  To study things like non-linear
variability properties \citep{Scaringi2012}, time series with
lengths of many days and good cadence are needed.  Kepler has provided
these for a handful of sources, while the Evryscope should be able
to provide such light curves for many more sources.  Additionally,
Kepler has provided such light curves for a highly biased sample of
cataclysmic variables -- the objects which are known to be bright CVs
ahead of time.  Transients with low duty cycles are not included in
the sample, so comparisons of their behaviour with that of the
persistently bright objects cannot be made.

%\begin{figure}
%\includegraphics[width=8 cm]{fig_scaringi.eps}
%\caption{Power spectra of the cataclysmic variable MV Lyrae.  Each power spectrum represents several days' to several weeks' worth of data, but it is also clear that there are features on timescales of a few minutes.  Probing this type of system can be done only with a high cadence, high duty cycle observatory.  Probing many of these systems can be done only with a project like {\it Evryscope}. Figure taken from Scaringi et al. 2012.}
%\end{figure}

X-ray binaries are another class of object which show anecdotal
evidence for minute-scale (and faster) optical variability, but which
can often be hard to study.  X-ray binary outbursts typically evolve
on timescales of weeks, meaning that it is difficult to obtain
sufficient target-of-opportunity time to sample them well, but it is
also not possible to study their bright phases with classically
scheduled planned observations.  It is clear that dramatic mid-IR
variability can be seen in X-ray binaries on fast timescales \citep{Gandhi2011}, and that strong variability on sub-second timescales can
be seen as well \citep{Kanbach2001}.  Having constant optical
monitoring to compare with intensive or all-sky monitoring
observations in the X-rays will provide a valuable resource.
Additionally, Type I X-ray bursts should be optically detectable from many
accreting neutron stars (see e.g. \citealt{Pedersen1982}), meaning that
the Evryscope will provide better monitoring of the rates of Type I
bursts for many more sources than X-ray monitors can provide.

\subsubsection{Spin-up of magnetic white dwarfs}
In a subset of accreting white dwarf systems, the magnetic field of
the white dwarf is strong enough to channel the accretion flow down
the magnetic pole.  When these systems have magnetic and rotation axes
for the white dwarf which are different, the emission varies
periodically on the spin period of the white dwarf due to a
``lighthouse effect''.  The typical spin periods of intermediate polars
are a few hundred seconds, well matched to the Evryscope's cadence,
so that with high cadence coverage, it should be possible to measure
their spin period evolution. We estimate that the rotation periods of 10-15 known sources will be observable with the Evryscope, and periodic emission may also be visible from a similarly-sized group of currently unknown objects. Testing whether the spin-up of magnetic
white dwarfs agrees with theoretical models will give a relatively
easy way to test the general theory of accretion torquing that is
often applied to explain how millisecond pulsars form in binaries with
neutron stars (e.g. \citealt{Smarr1976}).

\subsection{Unexpected stellar events}
\label{sec:unexpected_stellar}
Monitoring very large numbers of stars will nearly inevitably lead to discovering examples of very rare and/or unknown stellar variability. Long-term variable star monitoring has revealed truly spectacular objects and events -- for example, the V1309 Sco stellar merger \citep{Tylenda2011}. Detected as an eclipsing binary by the OGLE survey, the object showed a rapid period decrease followed by a very large outburst. After the outburst (unfortunately only noticed after the event) the object was no longer an eclipsing binary, displaying the light curve and spectrum of only a single star (albeit rather a disturbed one). It is likely that there are similar objects waiting to coalesce elsewhere in the galaxy; monitoring very large samples of eclipsing binaries with Evryscope systems could reveal dangerously decreasing periods well before the events. 

Similarly, drop-out events produced by eclipsing dust disks should be easy to detect as disappearing or dimming stars in the Evryscope survey, especially in star-formation regions, where monitoring only $\sim$10,000 targets could produce several dust-disk eclipse events on the timescales of the Evryscope survey \citep{Mamajek2012, Kenworthy2014}. High-time-resolution, high-precision eclipse light curve structures of these objects could be used to constrain the small-scale structure of dust disks which lead to planetary systems \citep{Rodriguez2013}.

\section{Faint-object regime: Rare, bright extragalactic transients}
\label{sec:extragalactic}
The first near-continuous minute-by-minute record of everything that occurs in very large fractions of the sky will include both known objects and unexpected transient events. For the first time in the optical, transient events will be detected and recorded wherever they occur on minute timescales, without requiring telescope pointing. Since all the data will be recorded, it will be possible to achieve deep co-adds for the detection of extragalactic transients; the Evryscope prototype will achieve a per-hour limiting magnitude of $\rm{m_V}$=18.2 in sky regions with median crowding.

\begin{deluxetable*}{llllllll}
\tablecaption{\label{tab:current_sne}The Evryscope compared to current multiple-telescope extragalactic transient surveys}
\tabletypesize{\footnotesize}
\startdata
\bf Survey & \bf FOV / sq. deg. & \bf Aperture / mm & \bf \arcsec/pixel & \bf Sites & \bf Pixels / site & \bf Targets & \bf Ref.\\

ATLAS & 30 & 500 & 1.9 & 2 & 100 MPix & NEOs \& general & \citet{Tonry2011} \\
ROTSE-III & 3.4 & 450 & 3 & 4 (4 tels.) & 16MPix & GRBs & \citet{Yost2006}\tablenote{Multiple sites for continuous observation}\\
ASAS-SN& 162 & 140 & 7 & 2 (8 tels.) & 16 MPix & Supernovae & \citet{Shappee2014}\tablenote{Multiple sites observing different fields}\\
TAROT & 3.4 & 250 & 3.3 & 2 & 8 MPix & GRBs & \citet{Klotz2009}\\
PROMPT & 0.18 & 410 & 0.6 & 1 (6 tels.) & 24 MPix & GRBs & \citet{Reichart2005}\\
Pi of the Sky & 7700 & 71 & 36 & 2 & 64 Mpix & GRBs (1d storage) & \citet{Malek2010} \\
Evryscope & 8,660 & 61 & 13.6 & 1 & 780 MPix & General & This paper\\
\end{deluxetable*}

In Table \ref{tab:current_sne} we compare the Evryscope's specifications to a selection of current small-aperture extragalactic transient surveys, showing the different parameter space the Evryscope concept addresses. We note that Pi of the Sky \citep{Malek2010} has Evryscope-like field of view, albeit with course pixels. However, it is a specialized short-cadence GRB survey with relatively large pixels and which does not record data long-term, and so it cannot perform the science cases described here.

\subsection{Nearby, Young Supernovae}
\label{sec:nearby_sne}
\begin{figure}
  \centering
  \resizebox{0.9\columnwidth}{!}
   {
   \includegraphics[]{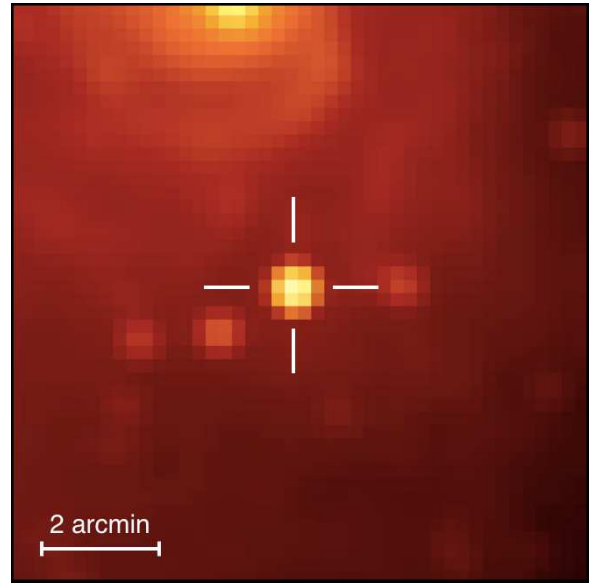}

   }
   \caption{A simulated Evryscope image of SN2011fe, a very nearby bright supernova detected by the Palomar Transient Factory \cite{Law09}. Based on an LCOGT image of the supernova soon after discovery \citep{Nugent2011}, with simulated 13\arcsec per pixel sampling and camera-lens point-spread-functions applied. The scale bar is 2 arcminutes long. The supernova is clearly distinguishable from the background structure of even this nearby, bright galaxy.}

   \label{fig:sn2011fe}
\end{figure}

While recent wide-field optical surveys such as PTF, Pan-STARRS, and CRTS have discovered supernovae at a remarkable rate (PTF alone has spectroscopically confirmed 2000), by far the highest impact discoveries have resulted from the most nearby events (distance $\lsim$20 Mpc). The detailed multi-wavelength studies enabled for these sources more than compensates for their relative rarity. The most prominent example is SN2011fe in M101, which for the first time provided direct evidence of a white dwarf progenitor for a type Ia supernova \citep{Nugent2011, Li2011, Bloom2012}.  Because of its continuous coverage, the Evryscope offers a promising avenue for discovery of such nearby supernovae, with the extremely valuable addition that such sources will be discovered when they are extremely young.  By probing the regime of shock breakout (i.e., when the optical depth drops below unity as the outgoing blast wave breaks outs of the stellar envelope), the Evryscope will enable direct measurements of the progenitor radii of the few nearby supernovae it discovers each year \citep{Rabinak2011, Kasen2010, Piro2010}.  With the Evryscope's pre-discovery imaging it will be possible to search for pre-explosion outbursts. These outbursts have been observed for a number of type IIn supernova (e.g., \citealt{Pastorello2007, Ofek2013}) and are a sensitive probe of mass-loss in the final stages of the evolution of massive stars. Despite the relatively low angular resolution of currently-feasible systems, most nearby supernovae will be easily distinguished from their host galaxies (e.g. Figure \ref{fig:sn2011fe}).

\subsection{Gamma-Ray Bursts}
\label{sec:GRBs}
The most luminous known class of extragalactic transients are the afterglows of gamma-ray bursts (GRBs).  Since the launch of the Swift satellite in 2004, a large number of robotic optical telescopes were built to automatically respond to GRB triggers as promptly as possible, in some cases capturing associated optical emission while the prompt gamma-ray emission was still ongoing (e.g., \citealt{Akerlof1999, Vestrand2005, Racusin2008}).  The nature of this ``prompt'' optical emission, and its relation to the high-energy emission, remains controversial.  The Evryscope will push beyond this paradigm to generate continuous imaging of GRB fields on minute time scales, not only immediately following GRB triggers, but pre-imaging those fields on minute cadences before the explosion.  Using the all-sky rate of Swift GRBs ($\sim$630 per year), together with the observed optical luminosity function \citep{Cenko2009}, we estimate the Evryscope will be capable of detecting $\sim$10 GRB afterglows each year, each with exquisite sampling on minute time scales, allowing for detailed comparison with high-energy (gamma and X-ray) light curves.  

\subsection{Gravitational wave electromagnetic counterparts}
\label{sec:ligo}
The improving sensitivity of gravitational wave searches is leading to efforts to prepare for simultaneous electromagnetic detection of gravitational wave triggers (e.g. \citealt{Kasliwal2014}, \citealt{LIGO2013}, LSC13 hereafter). Although the optical properties of the gravitational wave sources are quite uncertain, current sky surveys such as PTF, ROTSE, Skymapper, etc. are capable of setting useful limits on a gravitational wave source population – as exemplified by the initial searches and preparations detailed in LSC13. These searches, however, rely on the telescopes receiving a timely gravitational wave trigger signal, a very challenging proposition because it requires a realtime GW-detection pipeline and rapid telescope pointing. Even worse, the localizations from LIGO are uncertain at the degrees or tens-of-degrees level, necessitating rapid optical follow-up to reduce the number of variable and transient candidates that could be coincident with the gravitational wave. The Evryscope has a similar or better per-minute limiting magnitude than four of the nine LSC13 searches, and it has a snapshot coverage thousands of times larger than any that approach its sensitivity. The standard Evryscope observing mode will allow after-the-fact detection of gravitational-wave associated transients, without any need for rapid triggers. This capability will enable a search for optical counterparts to gravitational waves pulled out of the noise even years after the data was taken.

\subsection{Unknown or unexpected transients}
\label{sec:unknown_transients}
The Evryscope's very rapid cadence, extremely large field of view, and large \'etendue explores a new region of survey parameter space. As such, it is possible that the survey will reveal new unknown optical transients that would be rejected as cosmic rays or single-detection asteroids in longer-cadence surveys. For example: extremely fast radio transients with currently unknown origins have recently been discovered \citep{Thornton2013, Lorimer2013, Trott2013, Coenen2014}. Due to their rarity and millisecond-scale speed, there is currently no way to get useful constraints on their optical brightness. The Evryscope dataset will allow us to obtain simultaneous optical brightness limits (or even detections) on a minute-by-minute basis – without any need for triggering or pointing at these or similar targets. This mode will allow confirmation of transients detected in archival data taken at other wavelengths (e.g. \citealt{Law2004}), or in new all-sky surveys such as the Owens Valley LWA \citep{Hallinan2014} or LOFAR \citep{Haarlen2013}.

\section{Summary}
\label{sec:summary}
The all-sky gigapixel-scale telescope concept offers the possibility of opening a new parameter space for optical astronomical telescopes, where every possible target is observed simultaneously, whenever the sky is dark. Because they integrate for hours each night on every part of the sky, these systems can collect competitive numbers of photons per night compared to larger telescopes with smaller fields of view which can only observe each part of the sky for minutes.

Evryscopes will be able to contribute to exoplanet science using the uniquely wide field of view, including transiting planets around stars that are rare across the sky (for example, bright stars, nearby stars, bright M-dwarfs, bright white-dwarfs); monitoring large populations of eclipsing binaries for eclipse timing variations induced by planets; and monitoring large samples of nearby stars for microlensing events. Their surveys will also be complementary to more-targeted surveys such as TESS, providing the long-time-baseline required to characterize targets and find longer-period giant planets.

By simultaneously monitoring many millions of stars at high cadence, Evryscopes will be able to find young stars via their variability, greatly increase the known numbers of long-period eclipsing binaries that will constrain the stellar mass-radius-relation, explore the physics of white dwarf pulsations and accreting sources, and search for rare stellar merger and dust eclipse events.

In extragalactic science, near-future Evryscopes can detect gamma-ray-bursts as they go off, and have the capability of monitoring young nearby supernovae as they happen. Because they will generate a continuous movie of the sky, they have the capability to ``pre-image'' transient events, searching for outbursts on minute to year timescales before the events themselves occur. 

Current and near-future Evryscopes are limited by their relatively large pixels which limit their ability to detect faint targets, precluding their use for moderate and high-redshift transients, but as consumer imaging technology and computing capabilities improves this limit will be continuously improved. Further improvements to the systems' sky coverage by deploying next-generation systems to multiple observing sites, including Polar locations (e.g. \citealt{Law2013}), have the potential to produce the first deep and rapid all-the-sky, all-the-time synoptic survey.

\acknowledgments
\section*{Acknowledgements}
We thank the referee for comments which significantly improved the paper. We also thank Raymond Carlberg, Rick Murowinski and Suresh Sivanandam for interesting discussions during the conceptual design of the Arctic Evryscope.  This research was supported by the NSF grant AST-1407589.  
\\

\end{document}